\PassOptionsToPackage{unicode}{hyperref}
\PassOptionsToPackage{hyphens}{url}
\PassOptionsToPackage{dvipsnames,svgnames,x11names}{xcolor}
\documentclass[
]{article}

\usepackage{amsmath,amssymb}
\usepackage{iftex}
\ifPDFTeX
  \usepackage[T1]{fontenc}
  \usepackage[utf8]{inputenc}
  \usepackage{textcomp} % provide euro and other symbols
\else % if luatex or xetex
  \usepackage{unicode-math}
  \defaultfontfeatures{Scale=MatchLowercase}
  \defaultfontfeatures[\rmfamily]{Ligatures=TeX,Scale=1}
\fi
\usepackage{lmodern}
\ifPDFTeX\else  
\fi
\IfFileExists{upquote.sty}{\usepackage{upquote}}{}
\IfFileExists{microtype.sty}{% use microtype if available
  \usepackage[]{microtype}
  \UseMicrotypeSet[protrusion]{basicmath} % disable protrusion for tt fonts
}{}
\makeatletter
\@ifundefined{KOMAClassName}{% if non-KOMA class
  \IfFileExists{parskip.sty}{%
    \usepackage{parskip}
  }{% else
    \setlength{\parindent}{0pt}
    \setlength{\parskip}{6pt plus 2pt minus 1pt}}
}{% if KOMA class
  \KOMAoptions{parskip=half}}
\makeatother
\usepackage{xcolor}
\ifx\paragraph\undefined\else
  \let\oldparagraph\paragraph
  \renewcommand{\paragraph}[1]{\oldparagraph{#1}\mbox{}}
\fi
\ifx\subparagraph\undefined\else
  \let\oldsubparagraph\subparagraph
  \renewcommand{\subparagraph}[1]{\oldsubparagraph{#1}\mbox{}}
\fi

\usepackage{color}
\usepackage{fancyvrb}

\DefineVerbatimEnvironment{Highlighting}{Verbatim}{commandchars=\\\{\}}
\usepackage{framed}
\definecolor{shadecolor}{RGB}{241,243,245}
\newenvironment{Shaded}{\begin{snugshade}}{\end{snugshade}}

\newcommand{\AttributeTok}[1]{\textcolor[rgb]{0.40,0.45,0.13}{#1}}

\newcommand{\CommentTok}[1]{\textcolor[rgb]{0.37,0.37,0.37}{#1}}

\newcommand{\DecValTok}[1]{\textcolor[rgb]{0.68,0.00,0.00}{#1}}

\newcommand{\FloatTok}[1]{\textcolor[rgb]{0.68,0.00,0.00}{#1}}
\newcommand{\FunctionTok}[1]{\textcolor[rgb]{0.28,0.35,0.67}{#1}}

\newcommand{\NormalTok}[1]{\textcolor[rgb]{0.00,0.23,0.31}{#1}}

\newcommand{\OtherTok}[1]{\textcolor[rgb]{0.00,0.23,0.31}{#1}}

\newcommand{\SpecialCharTok}[1]{\textcolor[rgb]{0.37,0.37,0.37}{#1}}

\newcommand{\StringTok}[1]{\textcolor[rgb]{0.13,0.47,0.30}{#1}}

\providecommand{\tightlist}{%
  \setlength{\itemsep}{0pt}\setlength{\parskip}{0pt}}\usepackage{longtable,booktabs,array}
\usepackage{calc} % for calculating minipage widths
\usepackage{etoolbox}
\makeatletter
\patchcmd\longtable{\par}{\if@noskipsec\mbox{}\fi\par}{}{}
\makeatother
\IfFileExists{footnotehyper.sty}{\usepackage{footnotehyper}}{\usepackage{footnote}}
\makesavenoteenv{longtable}
\usepackage{graphicx}
\makeatletter
\def\maxwidth{\ifdim\Gin@nat@width>\linewidth\linewidth\else\Gin@nat@width\fi}
\def\maxheight{\ifdim\Gin@nat@height>\textheight\textheight\else\Gin@nat@height\fi}
\makeatother
\setkeys{Gin}{width=\maxwidth,height=\maxheight,keepaspectratio}
\makeatletter
\def\fps@figure{htbp}
\makeatother
\newlength{\cslhangindent}
\newlength{\csllabelwidth}
\newlength{\cslentryspacingunit} % times entry-spacing
\newenvironment{CSLReferences}[2] % #1 hanging-ident, #2 entry spacing
 {% don't indent paragraphs
  \setlength{\parindent}{0pt}
  \ifodd #1
  \let\oldpar\par
  \def\par{\hangindent=\cslhangindent\oldpar}
  \fi
  \setlength{\parskip}{#2\cslentryspacingunit}
 }%
 {}
\usepackage{calc}

\usepackage{arxiv}
\usepackage{orcidlink}
\usepackage{amsmath}
\usepackage[T1]{fontenc}
\usepackage{colortbl}
\usepackage{bera}
\usepackage{xcolor}
\usepackage{hyperref}
\usepackage[utf8]{inputenc}
\def\tightlist{}

\makeatletter
\@ifpackageloaded{caption}{}{\usepackage{caption}}
\AtBeginDocument{%
\ifdefined\contentsname
  \renewcommand*\contentsname{Table of contents}
\else
  \newcommand\contentsname{Table of contents}
\fi
\ifdefined\listfigurename
  \renewcommand*\listfigurename{List of Figures}
\else
  \newcommand\listfigurename{List of Figures}
\fi
\ifdefined\listtablename
  \renewcommand*\listtablename{List of Tables}
\else
  \newcommand\listtablename{List of Tables}
\fi
\ifdefined\figurename
  \renewcommand*\figurename{Figure}
\else
  \newcommand\figurename{Figure}
\fi
\ifdefined\tablename
  \renewcommand*\tablename{Table}
\else
  \newcommand\tablename{Table}
\fi
}
\@ifpackageloaded{float}{}{\usepackage{float}}
\floatstyle{ruled}
\@ifundefined{c@chapter}{\newfloat{codelisting}{h}{lop}}{\newfloat{codelisting}{h}{lop}[chapter]}
\floatname{codelisting}{Listing}

\makeatother
\makeatletter
\@ifpackageloaded{caption}{}{\usepackage{caption}}
\@ifpackageloaded{subcaption}{}{\usepackage{subcaption}}
\makeatother
\makeatletter
\@ifpackageloaded{tcolorbox}{}{\usepackage[skins,breakable]{tcolorbox}}
\makeatother
\makeatletter
\@ifundefined{shadecolor}{\definecolor{shadecolor}{rgb}{.97, .97, .97}}
\makeatother
\makeatletter
\@ifpackageloaded{tikz}{}{\usepackage{tikz}}
\makeatother
        \newcommand*\circled[1]{\tikz[baseline=(char.base)]{
          \node[shape=circle,draw,inner sep=1pt] (char) {{\scriptsize#1}};}}  
                  
\ifLuaTeX
  \usepackage{selnolig}  % disable illegal ligatures
\fi
\IfFileExists{bookmark.sty}{\usepackage{bookmark}}{\usepackage{hyperref}}
\IfFileExists{xurl.sty}{\usepackage{xurl}}{} % add URL line breaks if available
\hypersetup{
  pdftitle={Towards a unified language in experimental designs propagated by a software framework},
  pdfauthor={Emi Tanaka},
  pdfkeywords={grammar of experimental designs, design of
experiments, comparative experiments, interface design, grammarware},
  colorlinks=true,
  linkcolor={blue},
  filecolor={Maroon},
  citecolor={Blue},
  urlcolor={Blue},
  pdfcreator={LaTeX via pandoc}}

\newcommand{\runninghead}{A Preprint }
\title{Towards a unified language in experimental designs propagated by
a software framework}
\author{
\textbf{Emi Tanaka}~\orcidlink{0000-0002-1455-259X}\\Biological Data
Science Institute\\Australian National
University\\Canberra\\\href{mailto:emi.tanaka@anu.edu.au}{emi.tanaka@anu.edu.au}}
\date{}
\begin{document}
\maketitle
\begin{abstract}
Experiments require human decisions in the design process, which in turn
are reformulated and summarized as inputs into a system (computational
or otherwise) to generate the experimental design. I leverage this
system to promote a language of experimental designs by proposing a
novel computational framework, called ``the grammar of experimental
designs'', to specify experimental designs based on an object-oriented
programming system that declaratively encapsulates the experimental
structure. The framework aims to engage human cognition by building
experimental designs with modular functions that modify a targeted
singular element of the experimental design object. The syntax and
semantics of the framework are built upon consideration from multiple
perspectives. While the core framework is language-agnostic, the
framework is implemented in the \texttt{edibble} R-package. A range of
examples is shown to demonstrate the utility of the framework.
\end{abstract}
{\bfseries \emph Keywords}
\def\sep{\textbullet\ }
grammar of experimental designs \sep design of
experiments \sep comparative experiments \sep interface design \sep 
grammarware

\ifdefined\Shaded\renewenvironment{Shaded}{\begin{tcolorbox}[frame hidden, sharp corners, interior hidden, borderline west={3pt}{0pt}{shadecolor}, boxrule=0pt, breakable, enhanced]}{\end{tcolorbox}}\fi

\hypertarget{sec-intro}{%
\section{Introduction}\label{sec-intro}}

Experimental designs offer a rigorous data collection protocol that
seeks to achieve pre-defined objectives by imposing purposeful choices
and control over experimental variables. The process of deliberation on
the final experimental design is just as important, if not more, to
identify any potential issues that can be addressed prior to the
execution of the experiment. The experimental design literature,
however, is often \emph{product-oriented} rather than
\emph{process-oriented}; in other words, the focus is on the end product
(the validity or efficiency of the planned analysis for the final
experimental design; or algorithmic aspects to generate the design)
rather than the process to the final design. Similar sentiment dates
back from decades ago (as echoed in, for example, David M. Steinberg and
Hunter 1984b and its discussions in response) with recognition that
deriving the experimental context (e.g.~defining aims and selecting
experimental factors) and communication are important for experimental
planning in the real world.

The experimental aim and variables may initially be ill-defined and
require iterative refining. In constructing a valid and efficient
experimental design, the experimental context is invaluable (see for
examples, Bishop, Petersen, and Trayser 1982; Hahn 1984). However, this
context can be either lost in dialogue or understood implicitly, and
consequently, the full context is often not explicitly transcribed. The
downstream effect of not explicitly transcribing the context can be
large: misunderstanding of the context, loss of knowledge transfer,
inappropriate experimental designs rendering the collected data
meaningless for confirmatory analysis, or bad analysis that disregards
some significant experimental context (e.g.~prediction using a variable
that was used to derive the response). If anything, investing in a
carefully planned experiment will provide more value than an analysis
that attempts to scavenge meaning from a botched up experiment. The
experimental context, however, is often stripped away or of an
afterthought in many experimental design software systems (Tanaka and
Amaliah 2022) thereby providing less room for the users to dwell on
possible broader concerns in the experimental design. Such software
systems may be an artifact of viewing experimentation in terms of
abstract mathematical models, which has the benefits of allowing
recognition of common ground in distinct experiments (David M. Steinberg
and Hunter 1984b), but at the cost of losing the context.

No experiment is conducted without a person initiating the experiment.
Multiple people with different expertise are typically involved in
planning and executing an experiment but human communication is a
complex process, let alone interdisciplinary communication that
compounds the challenge in achieving a shared understanding (Winowiecki
et al. 2011). David M. Steinberg and Hunter (1984a) specifically calls
out the statisticians ``by working to improve their interpersonal skills
and by studying some of the literature by pschologists, anthropologists,
and others concerning the interplay between technical and cultural
change''. Communication strategies can be employed to form mutual
understandings, however, these are not strict requirements for
generating an experimental design and (for the better or for the worse)
communications are largely left to the autonomy of each individual. This
means that the process is subject to large variation that can ultimately
affect the final experimental design and critically, the relevance and
quality of the experimental data.

Coleman and Montgomery (1993) proposed a systematic approach for
organizing written documentation of plans for industrial experiments.
David M. Steinberg and Hunter (1984b) claimed that continually asking
questions about the theory underlying an experiment is important. These
practices, and in more general, writing documentation and seeking
alternative views, should be a routine practice in experiments (or any
data collection activity in fact). However, in the absence of extrinsic
motivation, we rely on individual's intrinsic motivation to adopt better
practices. Fishbach and Woolley (2022) proposed that the causes of the
intrinsic motivation are characterised by the direct association of the
activity and goal. In experimental design, our ultimate goal is to
collect experimental data that can be used as empirical evidence to
satisfy the experimental aim. This goal can be achieved without any of
the aforementioned practices. Consequently, better practices of
experimental design require the consideration of factors to increase the
motivation to adopt those practices.

The main contribution of this article is a computational framework for
constructing an experimental design based on a declarative system that
encapsulates experimental structures in a human-centered interface
design, with justification of the framework from multiple perspectives.
The core framework exposes the intermediate processes that make up the
final experimental design, using a cognitive approach that possibly
addresses some aforementioned challenges. Section~\ref{sec-background}
provides some background and defines terminology to explain the proposed
framework described in Section~\ref{sec-framework}.
Section~\ref{sec-examples} demonstrates the utility of the framework
using illustrative examples and Section~\ref{sec-discuss} concludes with
a discussion.

\hypertarget{sec-background}{%
\section{Background}\label{sec-background}}

In this section, I outline some concepts, many of which transcends the
field of experimental design that are relevant to the proposed framework
presented in Section~\ref{sec-framework}.

\hypertarget{sec-grammar}{%
\subsection{Grammarware}\label{sec-grammar}}

A \emph{grammar} combines a limited set of words under shared linguistic
rules to compose an unlimited number of proper sentences. In information
technology, computational objects governed by a set of processing rules
constitute a \emph{grammar}. Klint, Lämmel, and Verhoef (2005) coined
the term ``grammarware'' to refer to grammar and grammar-dependent
software from the perspective of engineering. Some examples of
grammarware used prominently in statistics are described next.

Wilkinson (2005) proposed the concept of ``the grammar of graphics'' as
an object-oriented graphics system that declaratively builds
quantitative graphics by specifying relatively modular components (data,
statistical transformation, scale, coordinate system, guide and
graphical layers that include information about graphical primitives and
mapping of data variables to aesthetic attributes), assembling a scene
from specifications stored as an object in a tree structure, and then
displaying it by translating the assembled object into a graphical
device. The most popular interpretation of the grammar of graphics is
the ggplot2 package (Wickham 2016) in the R language (R Core Team 2020),
but variations exist in other languages as well, such as Gadfly (Jones
et al. 2018) in Julia (Bezanson et al. 2017) and plotnine (Kibirige et
al. 2022) in Python (Van Rossum and Drake 2009). The realization of the
grammar of graphics aids users to flexibly build unlimited graphs from a
limited set of ``words'' (functions).

Another grammar is Structured Query Language (SQL), which is a
declarative language used to query and manipulate data. SQL statements
include special English keywords (e.g.~select, inner join, left join,
where, and group by) to specify the query in the identified database.
SQL statements can include nested queries such that the result of the
previous query is piped into the next query. A similar language was
employed in the dplyr package (Wickham et al. 2022) in R, referred to as
``the grammar of data manipulation'' by the authors. The core functions
in dplyr require both the first input and output to be objects of the
class data.frame (i.e., data in a tabular format), which allows
functions to be easily piped in a fashion similar to nested queries in
SQL. Each function is designed to perform a single task. The function
names correspond to English words, similar to the keywords in SQL.

The widespread use of these declarative languages is perhaps a testament
to the usefulness of these approaches. For more details and examples,
readers are recommended to look at the vignettes and documentation of
the packages.

\hypertarget{sec-comm}{%
\subsection{Communication Strategies}\label{sec-comm}}

An experiment is a human endeavour that generally involves more than one
person. Successfully completing an experiment typically hinges on the
communication between multiple people with their own expertise. Let us
consider a scenario where four actors are involved in an experiment:

\begin{itemize}
\tightlist
\item
  the \textbf{\emph{domain expert}} who drives the experimental
  objective and has the intricate knowledge of the subject area,
\item
  the \textbf{\emph{statistician}} who creates the experimental design
  layout after taking into account statistical and practical
  constraints,
\item
  the \textbf{\emph{technician}} who carries out the experiment and
  collects the data, and
\item
  the \textbf{\emph{analyst}} who analyses the experimental data and
  help interpret it.
\end{itemize}

The actors are purely illustrative and in practice, multiple people can
take on each role, one person can take on multiple roles, and a person
is not necessarily a specialist in the role assigned (e.g.~the role of
the \emph{statistician} can be carried out by a person whose primarily
training is not in statistics). The \emph{statistician} and
\emph{analyst} may be the same individual but the roles are explicitly
differentiated to signal that this is not always the case. All roles can
be performed by a single individual.

The scenario can begin with the \emph{domain expert} coming up with a
hypothesis or question and recruiting a \emph{statistician} to help
design the experiment. Before a \emph{statistician} can produce the
design layout, they must converse with the \emph{domain expert} to
understand the experimental objective, resources, practical constraints
and other possible nuances that might influence the outcome of the
experiment. There may be several communications before reaching a shared
understanding. The \emph{statistician} produces the final experimental
design along with an analysis plan. Once the design layout is produced,
these may be passed to a \emph{technician} to carry out the experiment
as per intended and collect the data. The \emph{analyst} then extracts
information, perhaps using the analysis plan by the \emph{statistician},
from the collected data with the help of the \emph{domain expert} for
the interpretation. Each actor plays a vital role in the experiment; if
even one actor fails in their role, then the whole experiment could be
in jeopardy, and in the worst case scenario, resources go to complete
waste. Even in this simple scenario, we can see that there are many
possible interactions between people with every chance of ``human
error'' in the communication.

How might we improve this interdisciplinary communication? Bracken and
Oughton (2006) highlighted the importance of language in
interdisciplinary research and insisted interdisciplinary projects must
allocate time to develop shared vocabularies. Winowiecki et al. (2011)
employed scenario building techniques as a tool for interdisciplinary
communication to promote structured dialogue to brainstorm particular
aspects of the problem. Ideally, we would like to employ a systematic
approach that abstracts the problem (and the solution) into a shared
understanding.

Not all experiments involve more than one person. In the special case
where only a single individual is involved, intra-personal communication
to internalize their experimental understanding must still take place,
and externalizing this understanding by transcribing or otherwise is
still important for the future self and others that wish to validate the
experimental data. Indeed, Nickerson (1999) conjectures reflection on
one's own knowledge and evaluation or justification of one's views as
some possible countermeasures to overimputing one's knowledge to others,
thus mitigating misunderstandings.

\hypertarget{sec-ed}{%
\subsection{Terminologies in Experimental Design}\label{sec-ed}}

The field of experimental design is large, and its domain application
(e.g., biology, psychology, marketing, and finance) also large. Numerous
terminologies are used to describe various aspects or components of the
experiment. Some terms apply only to particular domains; therefore,
their meaning is not commonly understood across domains; e.g.,
\emph{stimuli} are often treatments in behavioural science;
\emph{cluster} and \emph{block} can be used interchangeably -- the
former term is more likely used in clinical trials. Terms like
\emph{experimental unit} (smallest unit that independently receives the
treatment), \emph{observational unit} (smallest unit in which the
measurement is recorded on) and \emph{treatments} (a set of conditions
allocated to experimental units) are perhaps more universally
understood. In a comparative experiment, a \emph{control} usually refers
to the treatment level that may be the baseline for comparison with
other treatment levels (a \emph{placebo} is a common control in
pharmaceutical experiments). A \emph{replication} (of a treatment level)
typically refers to the number of times the treatment level is tested.
For an overview, see Bailey (2008), Lawson (2015), Montgomery (2020), or
other books on experimental design.

Some terms are used to describe a complete experimental design (e.g.,
\emph{randomised complete block design}, \emph{balanced incomplete block
design}, and \emph{split-plot design}) with limited parameters, such as
the number of treatments and replications. These ``named'' designs are
handy to succinctly describe the experimental structure, but it can
create a barrier to understanding the experimental structure if you are
unfamiliar with it (e.g.~do you know what a \emph{beehive design} is?
For those curious, see F. B. Martin 1973).

The \emph{experimental structure} can be divided into two main
substructures: the \emph{unit structure} and the \emph{treatment
structure}. The unit structure for a completely randomized design is
unstructured. A randomized complete block design has a unit structure in
which experimental units are nested within blocks. A \emph{factorial
design} is a design in which there is more than one set of treatment
factors, where the combination of the treatment levels across those
factors compose the whole set of treatments; in such a case, we say that
the treatment has a factorial structure. A \emph{fractional factorial
experiment} is an experiment in which only a subset of treatment factor
combinations is observed.

In industrial experiments, experimental factors are largely classified
into \emph{control} (or \emph{primary}) \emph{factor}, \emph{constant
factor}, and \emph{nuisance factor} (Coleman and Montgomery 1993; Viles
et al. 2008). The control factors here are equivalent to the treatment
factors. The constant factors are those that are maintained at the same
level throughout the experiment, and nuisance factors are those that
cannot be controlled. A \emph{run} typically refers to a complete
replicate of an experiment.

The terminology in experimental design is certainly diverse. The
presented terms thus far represent only a fraction of terms used. This
complicates any notion of building a ``unified language'' to form a
common understanding.

\hypertarget{sec-framework}{%
\section{The Grammar of Experimental Designs}\label{sec-framework}}

In an object-oriented programming (OOP) system, the objects are basic
(and relatively modular) components of the system that contain data and
code. The grammar of experimental designs, referred simply as ``the
grammar'' henceforth, is a computational framework that employs the OOP
system that considers experimental design as a working object that users
progressively build by encapsulating the experimental structure
declaratively by defining basic experimental components. This section
describes the external abstraction of the framework and its contrast to
other systems. The application of the grammar is shown in
Section~\ref{sec-examples}.

\hypertarget{components-of-the-grammar}{%
\subsection{Components of the Grammar}\label{components-of-the-grammar}}

As discussed in Section~\ref{sec-ed}, the terminology for experimental
design is diverse. In forming the grammar, we must formulate objects and
their methods such that they are relatively modular building blocks for
the final experimental design (see Section~\ref{sec-grammar} for other
grammarwares). The guiding principles for determining the components of
the grammar are that the terms have to be:

\begin{enumerate}
\def\labelenumi{\arabic{enumi}.}
\tightlist
\item
  meaningful to a diverse set of people,
\item
  reflective of fundamental actions, thoughts or factors in experiments,
  and
\item
  atomic (i.e., cannot be inferred from the composite of other terms).
\end{enumerate}

In the grammar, we describe terms fundamentally by considering every
categorised entity (physical or otherwise) that may be (directly or
indirectly) linked to the experimental unit to be a \emph{factor}. Every
factor in the system is assigned an \emph{explicit role} that is stored
as a class. The three primary roles of a factor, as defined in
Table~\ref{tbl-explicit-role}, are \textbf{treatment}, \textbf{unit} and
\textbf{record}. The treatment and unit are encoded as separate classes
as these are always semantically distinguished in a comparative
experiment. A nuisance (or uncontrollable) factor or any responses can
be encoded as a record class. Under the abstraction in
Table~\ref{tbl-explicit-role}, factors such as blocks, clusters,
experimental units, observational units, and experimental run are all
just units. Arguably, the small finite number of classes makes it easier
to form a shared understanding and limits the introduction of jargon.
The grammar uses the relational links between factors to infer other
roles of the factor as described next.

\hypertarget{tbl-explicit-role}{}
\begin{longtable}[]{@{}
  >{\raggedright\arraybackslash}p{(\columnwidth - 4\tabcolsep) * \real{0.1538}}
  >{\raggedright\arraybackslash}p{(\columnwidth - 4\tabcolsep) * \real{0.4231}}
  >{\raggedright\arraybackslash}p{(\columnwidth - 4\tabcolsep) * \real{0.4231}}@{}}
\caption{\label{tbl-explicit-role}Definition of explicit roles in the
grammar with some examples. The three roles are to some degree
characterised by the level of control by the
experimenter.}\tabularnewline
\toprule\noalign{}
\begin{minipage}[b]{\linewidth}\raggedright
Role/Class
\end{minipage} & \begin{minipage}[b]{\linewidth}\raggedright
Definition
\end{minipage} & \begin{minipage}[b]{\linewidth}\raggedright
Examples
\end{minipage} \\
\midrule\noalign{}
\endfirsthead
\toprule\noalign{}
\begin{minipage}[b]{\linewidth}\raggedright
Role/Class
\end{minipage} & \begin{minipage}[b]{\linewidth}\raggedright
Definition
\end{minipage} & \begin{minipage}[b]{\linewidth}\raggedright
Examples
\end{minipage} \\
\midrule\noalign{}
\endhead
\bottomrule\noalign{}
\endlastfoot
treatment & A factor that is of primary interest and under complete
control by the experimenter. & Vaccine in vaccine trials. Drug in
pharmaceutical experiments. Variety in plant improvement programs. \\
unit & Any categorised entity (physical or otherwise) that is under some
control by the experimenter. & Patient in clinical trials. Block in
glasshouse experiments. Time in longitudinal experiments. Spatial index
(e.g.~row and column) in crop field trials. \\
record & An observed or uncontrollable factor in the experiment. &
Responses from observational units. Traits like sex, gender, height,
age, and so on of an individual (note some of these may be used as a
blocking factor, therefore should be units in that instance). \\
\end{longtable}

The relationship between factors assigns an \emph{implicit role}; e.g.,
if a treatment factor is allocated to a plot factor, then the plot is an
experimental unit. The implicit roles are summarized in
Table~\ref{tbl-implicit-role}. Users are not required to be explicit
about the implicit roles, instead they are required to be explicit about
the relationships of factors.

\hypertarget{tbl-implicit-role}{}
\begin{longtable}[]{@{}
  >{\raggedright\arraybackslash}p{(\columnwidth - 6\tabcolsep) * \real{0.2500}}
  >{\raggedright\arraybackslash}p{(\columnwidth - 6\tabcolsep) * \real{0.2500}}
  >{\raggedright\arraybackslash}p{(\columnwidth - 6\tabcolsep) * \real{0.2500}}
  >{\raggedright\arraybackslash}p{(\columnwidth - 6\tabcolsep) * \real{0.2500}}@{}}
\caption{\label{tbl-implicit-role}Implicit roles based on the
relationship between factors.}\tabularnewline
\toprule\noalign{}
\begin{minipage}[b]{\linewidth}\raggedright
Explicit role of A
\end{minipage} & \begin{minipage}[b]{\linewidth}\raggedright
Explicit role of B
\end{minipage} & \begin{minipage}[b]{\linewidth}\raggedright
A --\textgreater{} B relationship
\end{minipage} & \begin{minipage}[b]{\linewidth}\raggedright
Implicit role for B
\end{minipage} \\
\midrule\noalign{}
\endfirsthead
\toprule\noalign{}
\begin{minipage}[b]{\linewidth}\raggedright
Explicit role of A
\end{minipage} & \begin{minipage}[b]{\linewidth}\raggedright
Explicit role of B
\end{minipage} & \begin{minipage}[b]{\linewidth}\raggedright
A --\textgreater{} B relationship
\end{minipage} & \begin{minipage}[b]{\linewidth}\raggedright
Implicit role for B
\end{minipage} \\
\midrule\noalign{}
\endhead
\bottomrule\noalign{}
\endlastfoot
unit & unit & B is nested in A & Nested unit \\
treatment & unit & B is applied to A & Experimental unit \\
record & unit & B is measured on A & Observational unit \\
\end{longtable}

\begin{figure}

{\centering \includegraphics{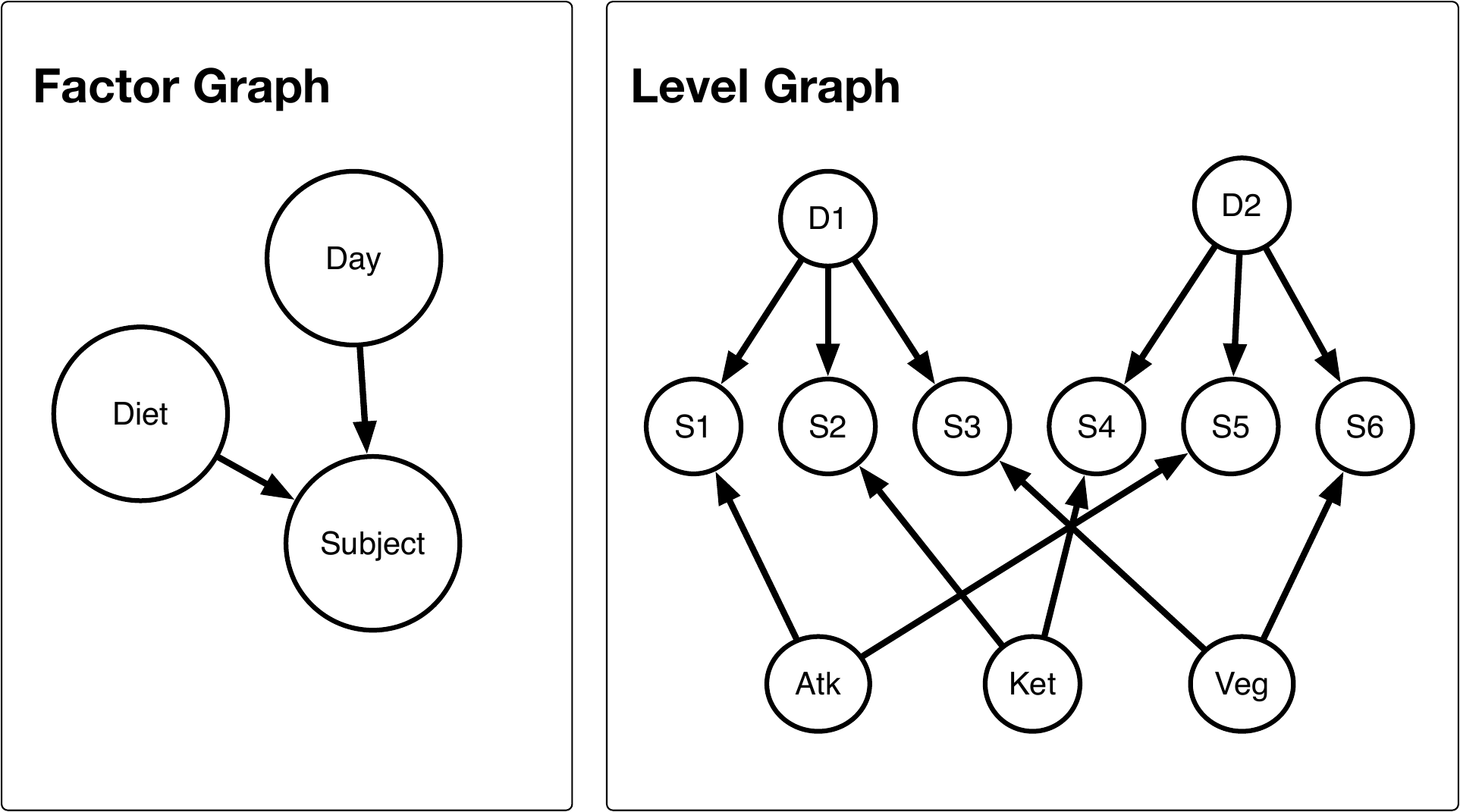}

}

\caption{\label{fig-logic}A visualisation of the graph form with three
factors: Diet, Subject, and Day. The experiment consists of three types
of diet (Atkins, Keto, and Vegan; labelled as the first three letters of
each diet) and three different subjects per day (labelled S1, S2, S3,
S4, S5, and S6) over a total of two days (labelled D1 and D2).}

\end{figure}

In the grammar, experimental designs are considered objects with two
forms: a graph form or a tabular form. The \textbf{graph form}
represents an intermediate construct of an experimental design as a pair
of directed acyclic graphs (DAGs) representing the high-level and the
low-level relationships (referred to as a factor graph and a level
graph, respectively). More specifically, in the \textbf{factor graph},
the nodes are factors and the edges are high-level relationships, while
in the \textbf{level graph}, the nodes are levels and the edges are the
low-level relationships. The direction of the edges specifies the
hierarchy between the nodes. An example of the graph form is shown in
Figure~\ref{fig-logic}.

The \textbf{tabular form} represents the final version of the
experimental design in a rectangular array where rows are the smallest
observational units and the columns are the variables or factors. This
tabular form, referred to as the \textbf{design table}, is a typical
output of an experimental design software.

\begin{figure}

{\centering \includegraphics{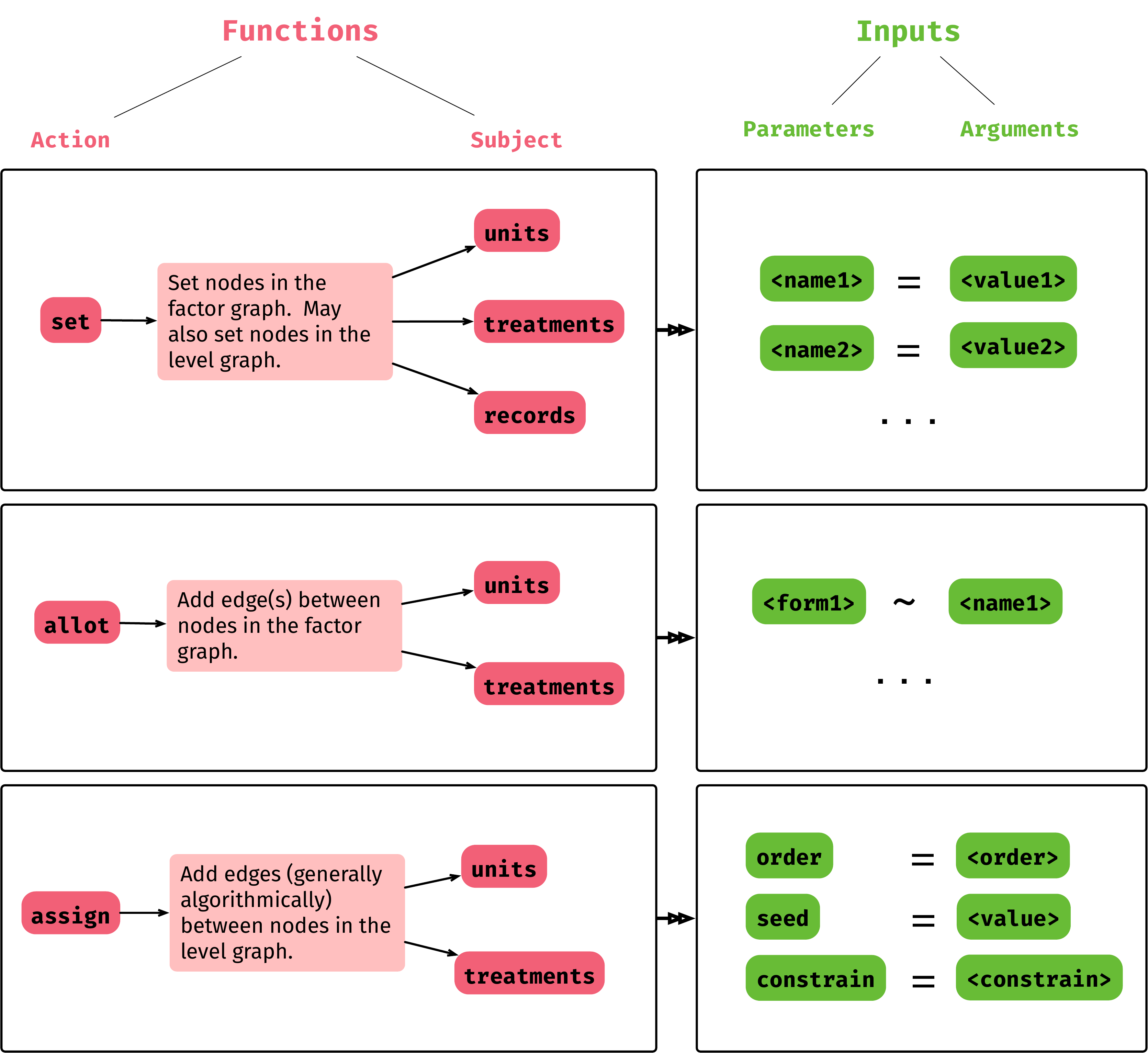}

}

\caption{\label{fig-lexicon}Visualisation of the lexicon of the concrete
syntax in the grammar of experimental designs. The name of the function
generally follows an action (verb) then the subject (noun). The
functions manipulate the graph form (e.g.~function \texttt{set\ units}
sets nodes with the unit class). The parameters
\texttt{\textless{}name1\textgreater{}} and
\texttt{\textless{}name2\textgreater{}} are user-defined factor names
(more than two can be defined as necessary) with the associated
arguments \texttt{\textless{}value1\textgreater{}} and
\texttt{\textless{}value2\textgreater{}} corresponding to the structure
of the factor (the number of levels, relationship with other factors,
etc). The parameter \texttt{\textless{}form1\textgreater{}} is specified
symbolically (e.g.~\texttt{trt1:trt2} is the combination of the
treatment factors \texttt{trt1} and \texttt{trt2}) alloted to the factor
\texttt{\textless{}name1\textgreater{}}. Again more allotment can be
specified as necessary. The argument
\texttt{\textless{}order\textgreater{}} is the algorithm that assigns
the treatments or units subject to the
\texttt{\textless{}constrain\textgreater{}} (typically the nesting
structure). The \texttt{\textless{}value\textgreater{}} for the
\texttt{seed} ensures that the design is reproducible.}

\end{figure}

The grammar begins with the initialization of the experimental design
object with an empty graph form. The user then declaratively manipulates
the object based on a small number of functions, as shown in
Figure~\ref{fig-lexicon}. The main actions are to either set the scene
(factors in the experiment), allot a factor to another factor, or assign
the levels to other levels algorithmically. The actions are concurrently
specified with the subject (primary roles); therefore, it is immediately
clear from the syntax which element of the experimental design object is
targetted. The actions, allot and assign, are made distinct as the
former is usually made explicit in dialogue and the latter is almost
always algorithmically derived. This concrete syntax may be altered
based on the domain specific language (as demonstrated later with the R
language in Section~\ref{sec-examples}). The object builds up
information on the experiment as the users specify the factors and their
relationships. When a user completes their specification, then they can
signal the conversion of the graph form to a tabular form. At this
stage, if the specification is valid (nodes in the level graph can all
be linked to one row), then it will render the design table.

It should be noted that not all experiments are comparative, i.e., some
experiments can have no treatment factors. The grammar does not require
specification of treatment factors although at the minimum requires
units to be specified.

\hypertarget{differences-to-other-systems}{%
\subsection{Differences to Other
Systems}\label{differences-to-other-systems}}

\begin{figure}

{\centering \includegraphics[width=0.8\textwidth,height=\textheight]{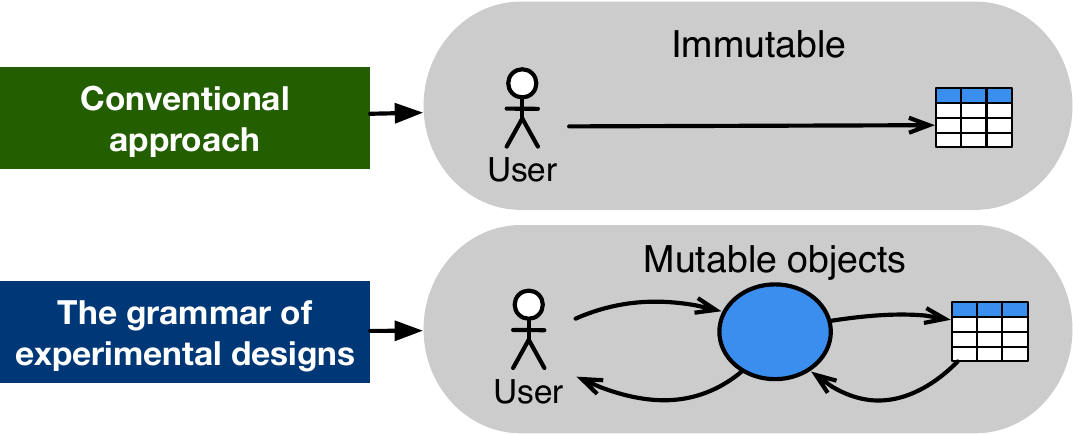}

}

\caption{\label{fig-mutable}The above figure depicts two pathways a user
may interact with a computational system. The bottom pathway allows
users to create mutable objects which maybe progressively modified. The
user receives feedback from objects that users may use for the next
step. In the top pathway, users create an immutable object which is
destroyed if an alternate output is seeked. In the top pathway,
user-object interaction is designed to be unidirectional and users are
not given opportunities to consciously think of intermediate processes.}

\end{figure}

By treating an experimental design as a mutable object, the grammar
allows a bi-directional interaction between the user and the object,
allowing users to inspect and progressively build the experimental
design. This bidirectional interaction is in contrast to many systems
that consider only unidirectional interactions, as illustrated in
Figure~\ref{fig-mutable}, where the major action of the user is to
specify a complete experimental design with no recourse to think about
individual components of the experiment.

\begin{figure}

{\centering \includegraphics{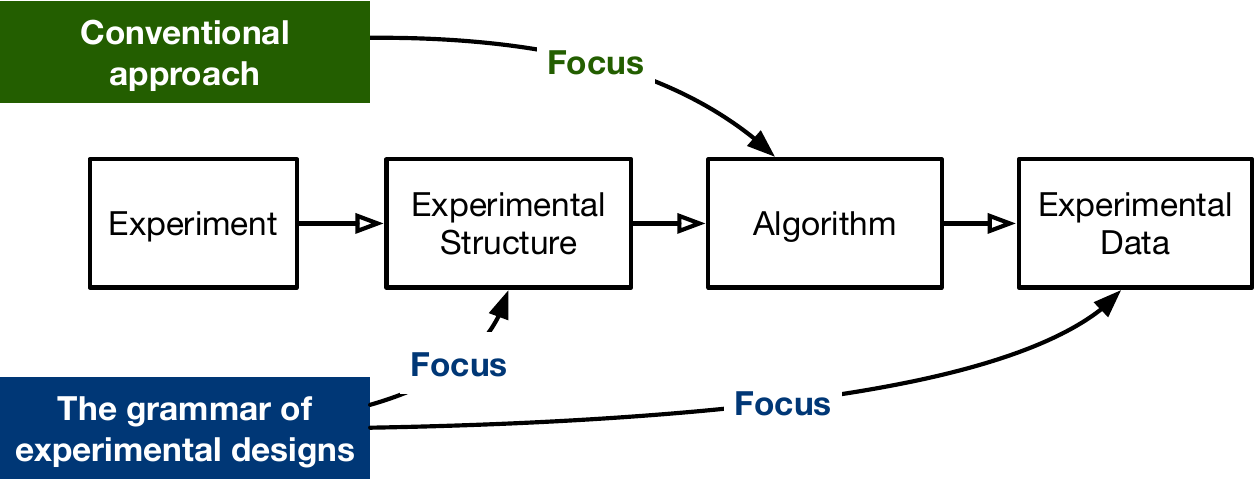}

}

\caption{\label{fig-opt}Every experiment has an experimental structure;
the experimental structure determines the algorithm for treatment
allocation; and the resulting allocation is used for collection of
experimental data. The conventional approach for the computational
generation of an experimental design is to focus on selecting an
algorithm with right inputs. The grammar shifts the focus to defining
the experimental structure and the desired output format. The approaches
are not mutually exclusive -- under the grammar, you can still select
the optimal algorithm.}

\end{figure}

Another key difference between the grammar and conventional approaches
for the computational generation of an experimental design, as
illustrated in Figure~\ref{fig-opt}, is that the grammar explicitly
defines the experimental structure and output. This does not mean that
the grammar cannot optimise the algorithmic assignment of the treatment
to units; the user can substitute the corresponding step as they see
fit. In this sense, the grammar is complementary to many existing
experimental design algorithms.

Furthermore, the grammar allows for various inputs that are fundamental
to experiments in a cognitive manner. In other words, the grammar treats
the specification of the experimental design as a structured dialogue.
Consider a scenario where a \emph{statistician} writes in their notes
during the meeting with the \emph{domain expert} where together they
decide on the structure of the experiment. Under the conventional
approach, when the \emph{statistician} enters the structure into the
computational system, the statistician has to reformulate this,
generally void of the context, to fit the system. By contrast, the
grammar is a more natural translation for the \emph{statistician} to map
their notes into the computational system. Indeed, the pre-design master
guide sheet by Coleman and Montgomery (1993) suggests a number of
elements (e.g.~response and treatment factors) that should be captured
in these notes that can be directly mapped in the grammar.

The example in Section~\ref{sec-classic} shows the difference in code
between the systems to specify the experimental design. While the code
is more verbose in the grammar, it should be clearer in communicating
the context of the experiment.

\hypertarget{sec-examples}{%
\section{Applications}\label{sec-examples}}

The grammar presented in Section~\ref{sec-framework} necessitates some
alterations when translated for a particular domain specific language.
For brevity, the translation of the grammar to the \texttt{edibble}
R-package (Tanaka 2023) in order to fit the particular nuances of the R
language and the user community is not described in this paper. This
section aims to demonstrate the utility of the grammar. Instructive
guide for the usage of the \texttt{edibble} R-package is reserved for
other avenues. The supplementary material shows the full design table
outputs and further explanations of the code.

In the following subsections, three examples of various flavours are
shown to illustrate the grammar of experimental designs described in
Section~\ref{sec-framework}. Section~\ref{sec-classic} demonstrates a
comparison of different programming approaches to achieve the same end
result. Section~\ref{sec-complex} deals with a complex nested design
showing how this can be specified using the grammar. Finally,
Section~\ref{sec-unbalanced} shows an example where the system can be
modified to deal with unbalanced cases.

\hypertarget{sec-classic}{%
\subsection{Classic Split-Plot Design}\label{sec-classic}}

\begin{figure}

{\centering \includegraphics{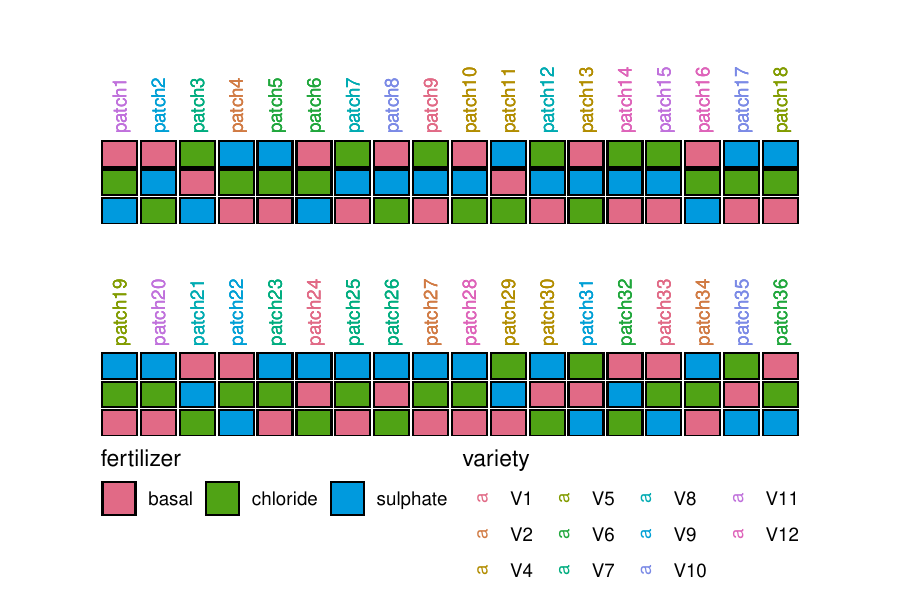}

}

\caption{\label{fig-exp1}A classic split-plot design with 36 patches
with 3 plots (depicted as a square) for each patch. The treatments are 3
levels of fertilizer (basal, chloride, and sulphase) and 12 levels of
variety. The color of the square indicate the fertilizer assigned to the
plot and the color of the text indicate which variety was assigned to a
patch.}

\end{figure}

Consider the classical split-plot experiment introduced by Fisher (1950)
where a land was divided into 36 patches, on which 12 varieties were
grown, and each variety planted in 3 randomly chosen patches. Each patch
was divided into three plots, with the plots randomly receiving either
the basal dressing only, sulphate or chloride of potash. In constructing
this experiment, the \emph{statistician} may have first randomized the
allocation of varieties to the patches with 3 replicates each and then
permuted the 3 fertilizer levels to the plots within each patch. A
random instance of this design is shown in Figure~\ref{fig-exp1}. The
original experiment measured the yield of each plot. Hypothetically, the
\emph{technician} may also record the biomass for each patch.

The construction of this design can follow in a procedural programming
manner where the 12 varieties with 3 replicates are permuted, followed
by replicating 36 times the permutation of 3 fertilizer levels. In the R
language, this may be coded like below. There may be further wrangling
to produce a design table.

\begin{Shaded}
\begin{Highlighting}[]
\NormalTok{variety }\OtherTok{\textless{}{-}} \FunctionTok{c}\NormalTok{(}\StringTok{"V1"}\NormalTok{, }\StringTok{"V2"}\NormalTok{, }\StringTok{"V3"}\NormalTok{, }\StringTok{"V4"}\NormalTok{, }\StringTok{"V5"}\NormalTok{, }\StringTok{"V6"}\NormalTok{, }
             \StringTok{"V7"}\NormalTok{, }\StringTok{"V8"}\NormalTok{, }\StringTok{"V9"}\NormalTok{, }\StringTok{"V10"}\NormalTok{, }\StringTok{"V11"}\NormalTok{, }\StringTok{"V12"}\NormalTok{)}
\NormalTok{fertilizer }\OtherTok{\textless{}{-}} \FunctionTok{c}\NormalTok{(}\StringTok{"basal"}\NormalTok{, }\StringTok{"sulphate"}\NormalTok{, }\StringTok{"chloride"}\NormalTok{)}
\FunctionTok{set.seed}\NormalTok{(}\DecValTok{1}\NormalTok{) }\CommentTok{\# for reproducibility}
\FunctionTok{sample}\NormalTok{(}\FunctionTok{rep}\NormalTok{(variety, }\AttributeTok{each =} \DecValTok{3}\NormalTok{)) }\CommentTok{\# variety allocation}
\FunctionTok{replicate}\NormalTok{(}\DecValTok{36}\NormalTok{, }\FunctionTok{sample}\NormalTok{(fertilizer)) }\CommentTok{\# fertilizer allocation}
\end{Highlighting}
\end{Shaded}

Alternatively, the structure of this design is well known as the
``split-plot design''. The \emph{statistician} may recognize the
structure to this ``named'' design, and generate this design via a
functional programming approach where the function name relates to the
name of the design. Below, we used the function \texttt{design.split()}
from the \texttt{agricolae} R-package (de Mendiburu 2021). Only two sets
of treatment factors are expected in a split-plot design, which is
reflected in the input parameter names \texttt{trt1} and \texttt{trt2}.
Notice that it is not immediately clear without further interrogation
which treatment factor is applied to the patches or the plots; in fact,
the units need not be defined.

\begin{Shaded}
\begin{Highlighting}[]
\NormalTok{agricolae}\SpecialCharTok{::}\FunctionTok{design.split}\NormalTok{(}\AttributeTok{trt1 =}\NormalTok{ variety,}
                        \AttributeTok{trt2 =}\NormalTok{ fertilizer,}
                        \AttributeTok{r =} \DecValTok{3}\NormalTok{, }
                        \AttributeTok{seed =} \DecValTok{1}\NormalTok{)}
\end{Highlighting}
\end{Shaded}

In the grammar, the design is progressively defined using a series of
composable operations as shown below.

\hypertarget{annotated-cell-3}{%
\label{annotated-cell-3}}%
\begin{Shaded}
\begin{Highlighting}[]
\FunctionTok{library}\NormalTok{(edibble)}
\NormalTok{des1 }\OtherTok{\textless{}{-}} \FunctionTok{design}\NormalTok{(}\StringTok{"Fisher\textquotesingle{}s split{-}plot design"}\NormalTok{) }\SpecialCharTok{\%\textgreater{}\%} \hspace*{\fill}\NormalTok{\circled{1}}
  \FunctionTok{set\_units}\NormalTok{(}\AttributeTok{patch =} \DecValTok{36}\NormalTok{, }\hspace*{\fill}\NormalTok{\circled{2}}
            \AttributeTok{plot =} \FunctionTok{nested\_in}\NormalTok{(patch, }\DecValTok{3}\NormalTok{)) }\SpecialCharTok{\%\textgreater{}\%}  
  \FunctionTok{set\_trts}\NormalTok{(}\AttributeTok{variety =} \DecValTok{12}\NormalTok{, }\hspace*{\fill}\NormalTok{\circled{3}}
           \AttributeTok{fertilizer =} \FunctionTok{c}\NormalTok{(}\StringTok{"basal"}\NormalTok{, }\StringTok{"sulphate"}\NormalTok{, }\StringTok{"chloride"}\NormalTok{)) }\SpecialCharTok{\%\textgreater{}\%}  
  \FunctionTok{set\_rcrds}\NormalTok{(}\AttributeTok{yield =}\NormalTok{ plot, }\hspace*{\fill}\NormalTok{\circled{4}}
            \AttributeTok{biomass =}\NormalTok{ patch) }\SpecialCharTok{\%\textgreater{}\%}  
  \FunctionTok{allot\_trts}\NormalTok{(variety }\SpecialCharTok{\textasciitilde{}}\NormalTok{ patch, }\hspace*{\fill}\NormalTok{\circled{5}}
\NormalTok{             fertilizer }\SpecialCharTok{\textasciitilde{}}\NormalTok{ plot) }\SpecialCharTok{\%\textgreater{}\%}  
  \FunctionTok{assign\_trts}\NormalTok{(}\AttributeTok{seed  =} \DecValTok{1}\NormalTok{,  }\hspace*{\fill}\NormalTok{\circled{6}}
              \AttributeTok{order =} \FunctionTok{c}\NormalTok{(}\StringTok{"random"}\NormalTok{, }\StringTok{"random"}\NormalTok{)) }\SpecialCharTok{\%\textgreater{}\%}   
  \FunctionTok{serve\_table}\NormalTok{() }\hspace*{\fill}\NormalTok{\circled{7}}
\end{Highlighting}
\end{Shaded}

\begin{description}
\tightlist
\item[\circled{1}]
The design object is initialised with an optional title of the
experiment.
\item[\circled{2}]
The units \texttt{patch} and \texttt{plot} are defined. The
\texttt{patch} has 36 levels while \texttt{plot} has 3 levels for each
\texttt{patch}.
\item[\circled{3}]
The treatments are \texttt{variety} with 12 levels and
\texttt{fertilizer} named as ``basal'', ``sulphate'' and ``chloride''.
\item[\circled{4}]
The records in the data collection will be \texttt{yield} for each
\texttt{plot} and the \texttt{biomass} for each \texttt{patch}.
\item[\circled{5}]
The treatments are allot to units. Specifically, \texttt{variety} to
\texttt{wholeplot} and \texttt{fertilizer} to \texttt{plot}.
\item[\circled{6}]
The treatments are then randomly assigned to corresponding unit
specified in the allotment. The \texttt{seed} is specified here so we
can replicate the results. It recognises that the \texttt{plot} is
nested in the \texttt{patch} and therefore uses this by default to
constrain the order that the treatment is allocated. Specifically, the
treatment order for both allotment are random.
\item[\circled{7}]
In the last step, we convert the intermediate design object into the
final experimental design table.
\end{description}

See Table 1 of the Supplementary Material for the full design table. The
Supplementary Material also shows the intermediate outputs and
explanation of other functions not shown here.

\hypertarget{sec-complex}{%
\subsection{Complex Nested Design}\label{sec-complex}}

\begin{figure}

{\centering \includegraphics{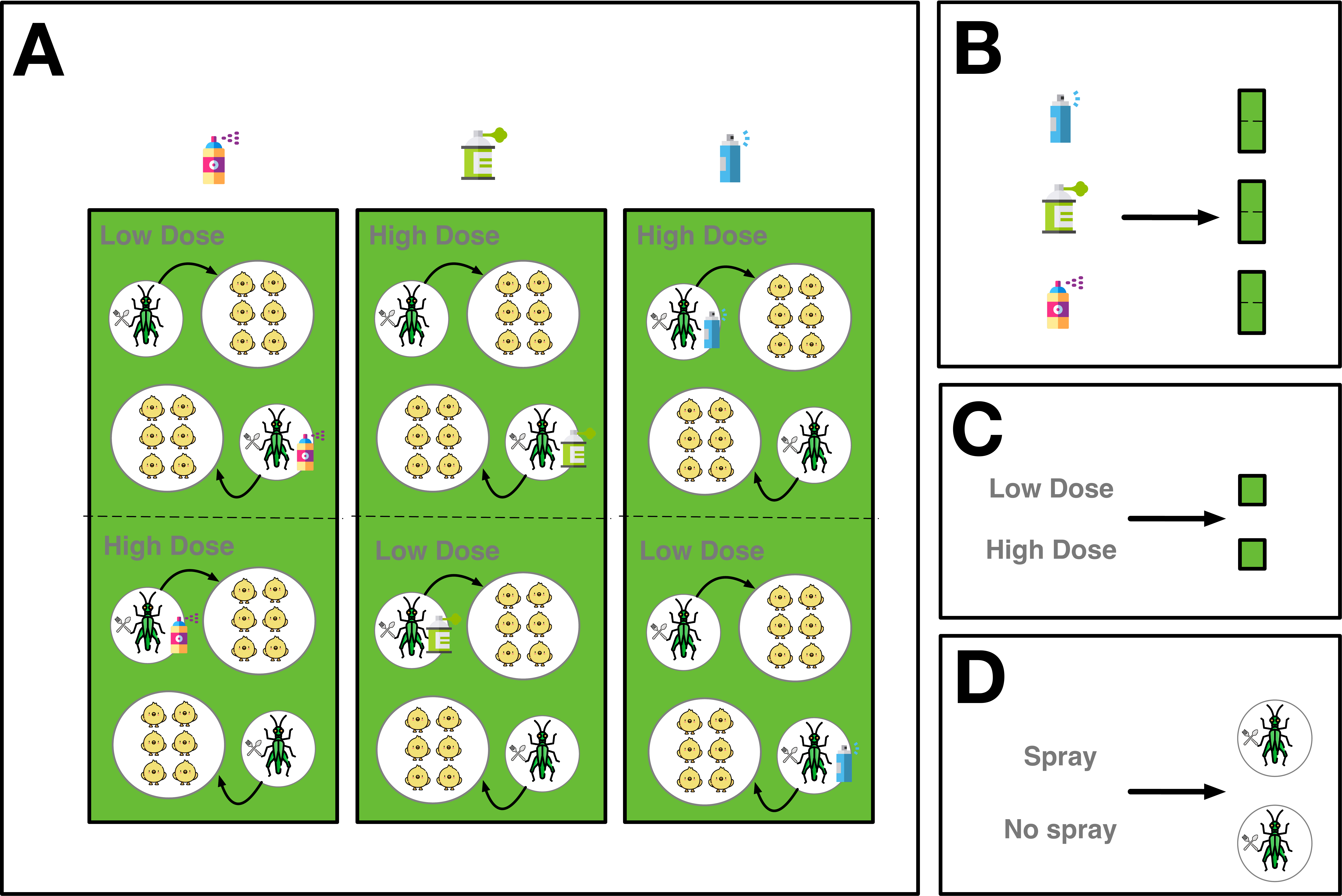}

}

\caption{\label{fig-exp2}A split-split plot design with three strips
that are divided into two swathes, which nests two pens with 6 chicks
each. The treatments were the three types of insecticide, the two dose
levels of the insecticide and whether the grasshoppers (food for the
chicks) were sprayed with insecticide or not. One insecticide was
randomly assigned to one strip. The dose of the insecticide was randomly
varied between the swath within a strip. The type of food (sprayed or
not) was randomly assigned to one pen within a swath. This experiment
was repeated each week for three weeks.}

\end{figure}

Consider next the experiment in P. A. Martin, Johnson, and Forsyth
(1996) aimed to investigate if insecticides used to control grasshoppers
affected the weight of young chicks of ring-necked pheasants, either by
affecting the grass around the chicks or by affecting the grasshoppers
eaten by the chicks. A description and illustration of the experiment is
in Figure~\ref{fig-exp2}.

Another random instance of the design in Figure~\ref{fig-exp2} is
specified in the grammar as follows.

\hypertarget{annotated-cell-4}{%
\label{annotated-cell-4}}%
\begin{Shaded}
\begin{Highlighting}[numbers=left,,]
\NormalTok{des2 }\OtherTok{\textless{}{-}} \FunctionTok{design}\NormalTok{(}\StringTok{"Complex nested factorial design"}\NormalTok{) }\SpecialCharTok{\%\textgreater{}\%} 
  \FunctionTok{set\_trts}\NormalTok{(}\AttributeTok{insecticide =} \DecValTok{3}\NormalTok{,  }\hspace*{\fill}\NormalTok{\circled{1}}
           \AttributeTok{dose\_level  =} \FunctionTok{c}\NormalTok{(}\StringTok{"low"}\NormalTok{, }\StringTok{"high"}\NormalTok{), }
           \AttributeTok{food\_type   =} \FunctionTok{c}\NormalTok{(}\StringTok{"sprayed"}\NormalTok{, }\StringTok{"unsprayed"}\NormalTok{)) }\SpecialCharTok{\%\textgreater{}\%}  
  \FunctionTok{set\_units}\NormalTok{(}\AttributeTok{week  =} \DecValTok{3}\NormalTok{, }\hspace*{\fill}\NormalTok{\circled{2}}
            \AttributeTok{strip =} \FunctionTok{nested\_in}\NormalTok{(week, }\DecValTok{3}\NormalTok{), }
            \AttributeTok{swath =} \FunctionTok{nested\_in}\NormalTok{(strip, }\DecValTok{2}\NormalTok{), }
            \AttributeTok{pen   =} \FunctionTok{nested\_in}\NormalTok{(swath, }\DecValTok{2}\NormalTok{), }
            \AttributeTok{chick =} \FunctionTok{nested\_in}\NormalTok{(pen, }\DecValTok{6}\NormalTok{)) }\SpecialCharTok{\%\textgreater{}\%} 
  \FunctionTok{allot\_trts}\NormalTok{(insecticide }\SpecialCharTok{\textasciitilde{}}\NormalTok{ strip, }\hspace*{\fill}\NormalTok{\circled{3}}
\NormalTok{              dose\_level }\SpecialCharTok{\textasciitilde{}}\NormalTok{ swath, }
\NormalTok{               food\_type }\SpecialCharTok{\textasciitilde{}}\NormalTok{ pen) }\SpecialCharTok{\%\textgreater{}\%}  
  \FunctionTok{assign\_trts}\NormalTok{(}\AttributeTok{seed =} \DecValTok{1}\NormalTok{) }\SpecialCharTok{\%\textgreater{}\%}  
  \FunctionTok{serve\_table}\NormalTok{()  }
\end{Highlighting}
\end{Shaded}

\begin{description}
\tightlist
\item[\circled{1}]
Here the treatment is defined first with 3 levels of insecticide, two
dose levels (low and high) and two food types (sprayed or unsprayed).
\item[\circled{2}]
The units are defined next. The experiment is run over 3 weeks. For each
week, there are 3 strips used. Each strip is split into two swathes.
Each swath has two pens. Each pen contains 6 chicks.
\item[\circled{3}]
Next we define the allotment of treatments to units. The insecticide is
alloted to strip, the dose level to swath and the food type to pen.
\end{description}

See Table 2 of the Supplementary Material for the full design table.

\hypertarget{sec-unbalanced}{%
\subsection{Unbalanced Factorial Design}\label{sec-unbalanced}}

Previous examples have conveniently used equal numbers of replicates for
each treatment, however, this is often not the case in practice. The
proposed system can cater for experiments with an unbalanced number of
treatments.

\begin{figure}

{\centering \includegraphics[width=0.5\textwidth,height=\textheight]{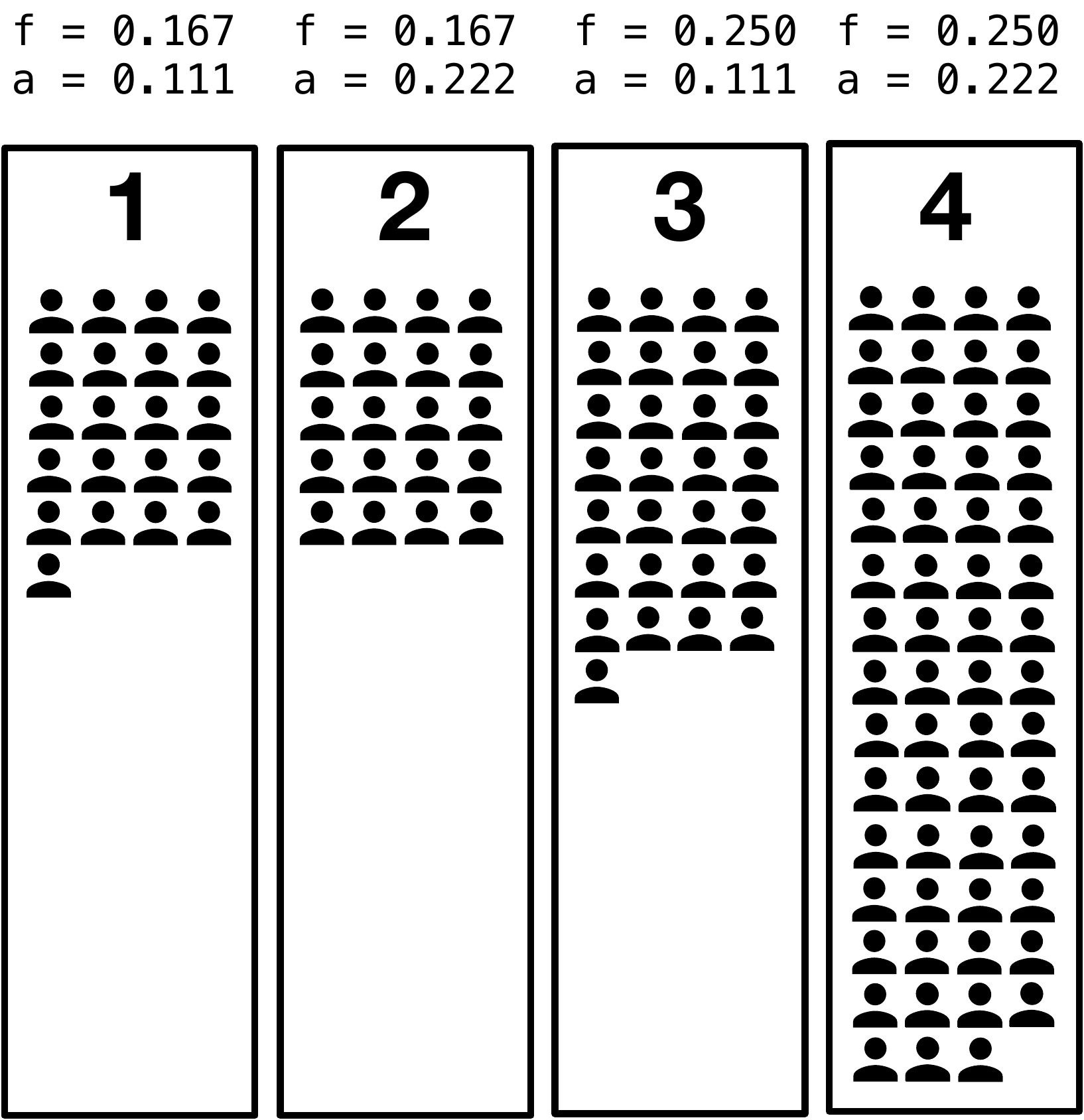}

}

\caption{\label{fig-exp3}There were four motion sickness experiments.
The first experiment consisted of 21 subjects with an acceleration (a)
of 0.111rms and frequency (f) of 0.167Hz. The next three consisted of
20, 29 and 59 subjects with a = 0.222 and f = 0.167, a = 0.111 and f =
0.250, and a = 0.222 and f = 0.250, respectively.}

\end{figure}

Suppose we consider the first four motion sickness experiments reported
by Burns (1984). The study, as shown in Figure~\ref{fig-exp3}, was a
collection of separate experiments. In this sense, the treatment
(acceleration and frequency) was pre-assigned and completely confounded
with the experiment.

This unbalanced design in Figure~\ref{fig-exp3} is specified in the
grammar as:

\hypertarget{annotated-cell-5}{%
\label{annotated-cell-5}}%
\begin{Shaded}
\begin{Highlighting}[numbers=left,,]
\NormalTok{des3 }\OtherTok{\textless{}{-}} \FunctionTok{design}\NormalTok{(}\StringTok{"Motion sickness incidence"}\NormalTok{) }\SpecialCharTok{\%\textgreater{}\%} 
  \FunctionTok{set\_units}\NormalTok{(}\AttributeTok{experiment =} \DecValTok{4}\NormalTok{,   }\hspace*{\fill}\NormalTok{\circled{1}}
            \AttributeTok{subject =} \FunctionTok{nested\_in}\NormalTok{(experiment,   }
                                \DecValTok{1} \SpecialCharTok{\textasciitilde{}} \DecValTok{21}\NormalTok{,  }
                                \DecValTok{2} \SpecialCharTok{\textasciitilde{}} \DecValTok{20}\NormalTok{,  }
                                \DecValTok{3} \SpecialCharTok{\textasciitilde{}} \DecValTok{29}\NormalTok{,  }
                                \DecValTok{4} \SpecialCharTok{\textasciitilde{}} \DecValTok{59}\NormalTok{)) }\SpecialCharTok{\%\textgreater{}\%}   
  \FunctionTok{set\_trts}\NormalTok{(}\AttributeTok{frequency    =} \FunctionTok{c}\NormalTok{(}\FloatTok{0.167}\NormalTok{, }\FloatTok{0.250}\NormalTok{),  }\hspace*{\fill}\NormalTok{\circled{2}}
           \AttributeTok{acceleration =} \FunctionTok{c}\NormalTok{(}\FloatTok{0.111}\NormalTok{, }\FloatTok{0.222}\NormalTok{)) }\SpecialCharTok{\%\textgreater{}\%}   
  \FunctionTok{allot\_trts}\NormalTok{(frequency}\SpecialCharTok{:}\NormalTok{acceleration }\SpecialCharTok{\textasciitilde{}}\NormalTok{ experiment) }\SpecialCharTok{\%\textgreater{}\%}   \hspace*{\fill}\NormalTok{\circled{3}}
  \FunctionTok{assign\_trts}\NormalTok{(}\AttributeTok{order =} \StringTok{"systematic"}\NormalTok{) }\SpecialCharTok{\%\textgreater{}\%}    \hspace*{\fill}\NormalTok{\circled{4}}
  \FunctionTok{serve\_table}\NormalTok{()}
\end{Highlighting}
\end{Shaded}

\begin{description}
\tightlist
\item[\circled{1}]
We specify that there are 4 experiments. Experiments 1, 2, 3 and 4 had
21, 20, 29 and 59 subjects, respectively.
\item[\circled{2}]
There were two treatment factors: frequency with two levels (0.167 and
0.250) and acceleration with two levels (0.111 and 0.222).
\item[\circled{3}]
The combination of the treatment factors are assigned to each
experiment.
\item[\circled{4}]
The allocation of the treatment is systematic.
\end{description}

See Table 3 of the Supplementary Material for the full design table.

\hypertarget{sec-discuss}{%
\section{Discussion}\label{sec-discuss}}

Multiple people with different expertise are typically involved in
planning and executing an experiment but communication is rarely easy or
seamless, especially across people from different domains. In designing
experiments, we ought to consider the time (Bracken and Oughton 2006)
and methods, such as structured dialogues (Winowiecki et al. 2011), to
form a shared understanding. A unified language in experimental designs
will aid in rapidly fostering mutual understanding among involved
parties. In this paper, I propose to leverage the design of the software
interface to promote a standardized grammar that govern the expression
of experimental designs in a structured approach.

A new framework, called ``the grammar of experimental designs'', was
presented as a process-based tool. The primary novel aspect of this
framework is that an experimental design is treated as a mutable object
that is progressively altered based on the explicit specifications of
fundamental experimental components. This approach exposes the
intermediate process to constructing the final experimental design, thus
providing a greater opportunity to notice any broader concerns in the
experimental designs. This in turn can encourage the investigation or
remedy of the experimental plan before its execution.

A number of functionalities are not discussed or demonstrated in this
paper in order to focus on the general framework rather than on specific
features. These functionalities include the specification of intended
observational records (including responses) of units; embedded data
validation for data entry; simulation of observational records;
diagnostics and visualization of designs. Abstract syntax and internal
object representation are also only briefly discussed. These
functionalities and internals warrant full discussion in separate
papers. Furthermore, an extended explanation of the \texttt{edibble}
package will be presented in other avenues. The framework does not
address all possible experimental structures but extensions of the
framework, such as situations with an undetermined number of levels or
complex conditional structures, can be envisioned as future research
directions.

This framework may be compelling for several reasons, some of which have
been outlined previously. First, explicit specification raises the
cognitive awareness of the experimental context and intention for both
the user and the reader. Second, it encourages encoding of information
as a machine-readable data, thereby allowing for further interrogation,
manipulation or even exportation to multiple formats. Third, it allows
for the partial specification of the experimental structure and permits
the reuse of the structure.

A recipe approach is often used for existing software to generate
randomized designs. A recipe or a named design describes an end product
and does not permit different processes to reach to a similar end
product. The grammar requires users to describe a particular course of
actions, thereby intentionally directing users to be explicit. This way
the software does not hinder the ability for users to encode more
information.

The proposed framework is purposefully designed such that it can be
extended and improved by other developers. For example, the assignment
of treatments (to units) can be substituted with alternative methods.
Arguably this step is the most algorithmically difficult part of the
process, and is the subject of many experimental design research. The
default assignment is currently simplistic. There will be many cases in
which the default system will not be suitable or is highly inefficient.
The goal of the grammar, however, is not to generate the most efficient
or optimal design for every experimental structure, which is an
impossible feat without user guidance. The goal of the grammar is to
standardize the specifications of the experimental structure so that we
can more easily form a shared understanding. As any other language, the
grammar of experimental designs has the potential to evolve.

In principle, the framework promotes good practice by requiring an
explicit specification of the elements of the experimental design.
However, principle alone is not sufficient to encourage mass adoption.
There are several possible extensions that make the framework attractive
despite its verbose specifications. These include immediate benefits
such as ease of adding data validation and automated visualization --
both of which are the subject of future papers. Fishbach and Woolley
(2022) suggested that immediate benefits can increase intrinsic
motivation. My hope is that these downstream features will eventuate in
the mass adoption of the framework, or even a similar framework, which
aids in the transparency of the experimental design process. We all gain
from better experimental practices. It is in this mass adoption, where
we come to share a unified language in experimental designs, that I
believe will aid in communication and result in the collective adoption
of better experimental designs. The practice of experimental design
requires holistic consideration of the total experimental process,
including that of psychological processes that translate to practice.

\hypertarget{supplementary-material}{%
\section{Supplementary Material}\label{supplementary-material}}

The supplementary material contains the full design table outputs from
the examples in Section~\ref{sec-examples} along with further
explanations of the code.

\hypertarget{acknowledgement}{%
\section*{Acknowledgement}\label{acknowledgement}}
\addcontentsline{toc}{section}{Acknowledgement}

This paper uses \texttt{knitr} (Xie 2015), \texttt{rmarkdown} (Xie,
Allaire, and Grolemund 2018) and Quarto (Posit 2023) for creating
reproducible documents. The code presented uses version 0.1.3 of the
\texttt{edibble} package available on CRAN. The latest development of
\texttt{edibble} can be found at
\url{https://github.com/emitanaka/edibble}.

\hypertarget{references}{%
\section*{References}\label{references}}
\addcontentsline{toc}{section}{References}

\hypertarget{refs}{}
\begin{CSLReferences}{1}{0}
\leavevmode\vadjust pre{\hypertarget{ref-Bailey2008-gw}{}}%
Bailey, Rosemary A. 2008. \emph{Design of {Comparative Experiments}}.
{Cambridge University Press}.

\leavevmode\vadjust pre{\hypertarget{ref-bezanson2017julia}{}}%
Bezanson, Jeff, Alan Edelman, Stefan Karpinski, and Viral B Shah. 2017.
{``Julia: A Fresh Approach to Numerical Computing.''} \emph{SIAM Review}
59 (1): 65--98. \url{https://doi.org/10.1137/141000671}.

\leavevmode\vadjust pre{\hypertarget{ref-bishopAnotherLookStatistician1982}{}}%
Bishop, Thomas, Bruce Petersen, and David Trayser. 1982. {``Another
{Look} at the {Statistician}'s {Role} in {Experimental Planning} and
{Design}.''} \emph{The American Statistician} 36 (4): 387--89.

\leavevmode\vadjust pre{\hypertarget{ref-Bracken2006-rk}{}}%
Bracken, L J, and E A Oughton. 2006. {``'What Do You Mean?' the
Importance of Language in Developing Interdisciplinary Research.''}
\emph{Transactions} 31 (3): 371--82.
\url{https://doi.org/10.1111/j.1475-5661.2006.00218.x}.

\leavevmode\vadjust pre{\hypertarget{ref-burnsMotionSicknessIncidence1984}{}}%
Burns, K. C. 1984.
{``\href{https://www.ncbi.nlm.nih.gov/pubmed/6466248}{Motion Sickness
Incidence: Distribution of Time to First Emesis and Comparison of Some
Complex Motion Conditions}.''} \emph{Aviation, Space, and Environmental
Medicine} 55 (6): 521--27.

\leavevmode\vadjust pre{\hypertarget{ref-colemanSystematicApproachPlanning1993a}{}}%
Coleman, David E, and Douglas C Montgomery. 1993. {``A {Systematic
Approach} to {Planning} for a {Designed Industrial Experiment}.''}
\emph{Technometrics} 35 (1): 1--12.

\leavevmode\vadjust pre{\hypertarget{ref-agricolae}{}}%
de Mendiburu, Felipe. 2021. \emph{Agricolae: Statistical Procedures for
Agricultural Research}.
\url{https://CRAN.R-project.org/package=agricolae}.

\leavevmode\vadjust pre{\hypertarget{ref-fishbachStructureIntrinsicMotivation2022}{}}%
Fishbach, Ayelet, and Kaitlin Woolley. 2022. {``The {Structure} of
{Intrinsic Motivation}.''} \emph{Annual Review of Organizational
Psychology and Organizational Behavior} 9 (1): 339--63.
\url{https://doi.org/10.1146/annurev-orgpsych-012420-091122}.

\leavevmode\vadjust pre{\hypertarget{ref-Fisher1950-hd}{}}%
Fisher, Ronald A. 1950. \emph{Statistical Methods for Research Workers}.
11th ed. {Oliver and Boyd}.

\leavevmode\vadjust pre{\hypertarget{ref-hahnExperimentalDesignComplex1984}{}}%
Hahn, Gerald J. 1984. {``Experimental {Design} in the {Complex
World}.''} \emph{Technometrics} 26 (1): 19--31.

\leavevmode\vadjust pre{\hypertarget{ref-daniel_c_jones_2018_1284282}{}}%
Jones, Daniel C., Ben Arthur, Tamas Nagy, Shashi Gowda, Godisemo, Tim
Holy, Andreas Noack, et al. 2018. {``GiovineItalia/Gadfly.jl: V0.7.0.''}
Zenodo. \url{https://doi.org/10.5281/zenodo.1284282}.

\leavevmode\vadjust pre{\hypertarget{ref-hassan_kibirige_2022_7124918}{}}%
Kibirige, Hassan, Greg Lamp, Jan Katins, gdowding, austin, Florian
Finkernagel, matthias-k, et al. 2022. {``{has2k1/plotnine: See the
{[}changelog{]}(https://plotn
ine.readthedocs.io/en/stable/changelog.html\#v0-9-0 ).}''} Zenodo.
\url{https://doi.org/10.5281/zenodo.7124918}.

\leavevmode\vadjust pre{\hypertarget{ref-Klint2005-iz}{}}%
Klint, Paul, Ralf Lämmel, and Chris Verhoef. 2005. {``Toward an
Engineering Discipline for Grammarware.''} \emph{ACM Transactions on
Software Engineering and Methodology} 14 (3): 331--80.
\url{https://doi.org/10.1145/1072997.1073000}.

\leavevmode\vadjust pre{\hypertarget{ref-Lawson2015}{}}%
Lawson, John. 2015. \emph{Design and {Analysis} of {Experiments} with
{R}}. {CRC Press}.

\leavevmode\vadjust pre{\hypertarget{ref-martinBeehiveDesignsObserving1973}{}}%
Martin, Frank B. 1973. {``Beehive {Designs} for {Observing Variety
Competition}.''} \emph{Biometrics} 29 (2): 397--402.
\url{https://doi.org/10.2307/2529404}.

\leavevmode\vadjust pre{\hypertarget{ref-martinEffectsGrasshoppercontrolInsecticides1996}{}}%
Martin, Pamela A., Daniel L. Johnson, and Douglas J. Forsyth. 1996.
{``Effects of Grasshopper-Control Insecticides on Survival and Brain
Acetylcholinesterase of Pheasant ( {\emph{Phasianus}}{ \emph{Colchicus}}
) Chicks.''} \emph{Environmental Toxicology and Chemistry / SETAC} 15
(4): 518--24.
\url{https://doi.org/10.1897/1551-5028(1996)015\%3C0518:EOGCIO\%3E2.3.CO;2}.

\leavevmode\vadjust pre{\hypertarget{ref-montgomeryDesignAnalysisExperiments2020}{}}%
Montgomery, D. 2020. \emph{Design and Analysis of Experiments}. 10th ed.
{Wiley}.

\leavevmode\vadjust pre{\hypertarget{ref-nickersonHowWeKnow1999}{}}%
Nickerson, Raymond S. 1999. {``How {We Know}\textemdash and {Sometimes
Misjudge}\textemdash{{What Others Know}}: {Imputing One}'s {Own
Knowledge} to {Others}.''} \emph{Psychological Bulletin} 125 (6):
737--59.

\leavevmode\vadjust pre{\hypertarget{ref-quarto}{}}%
Posit. 2023. \emph{Quarto: An Open-Source Scientific and Technical
Publishing System}. \url{https://quarto.org/}.

\leavevmode\vadjust pre{\hypertarget{ref-R-base}{}}%
R Core Team. 2020. \emph{R: A Language and Environment for Statistical
Computing}. Vienna, Austria: R Foundation for Statistical Computing.
\url{https://www.R-project.org/}.

\leavevmode\vadjust pre{\hypertarget{ref-steinbergExperimentalDesignReview1984a}{}}%
Steinberg, David M., and William G. Hunter. 1984a. {``{[}{Experimental
Design}: {Review} and {Comment}{]}: {Response}.''} \emph{Technometrics}
26 (2): 128. \url{https://doi.org/10.2307/1268106}.

\leavevmode\vadjust pre{\hypertarget{ref-steinbergExperimentalDesignReview1984}{}}%
Steinberg, David M, and William G Hunter. 1984b. {``Experimental
{Design}: {Review} and {Comment}.''} \emph{Technometrics} 26 (2):
71--97.

\leavevmode\vadjust pre{\hypertarget{ref-R-edibble}{}}%
Tanaka, Emi. 2023. \emph{{edibble}: Designing Comparative Experiments}.
\url{https://CRAN.R-project.org/package=edibble}.

\leavevmode\vadjust pre{\hypertarget{ref-Tanaka2022-hc}{}}%
Tanaka, Emi, and Dewi Amaliah. 2022. {``Current State and Prospects of
{R}-Packages for the Design of Experiments.''}
\url{https://doi.org/10.18637/jss.v096.i01}.

\leavevmode\vadjust pre{\hypertarget{ref-python}{}}%
Van Rossum, Guido, and Fred L. Drake. 2009. \emph{Python 3 Reference
Manual}. Scotts Valley, CA: CreateSpace.

\leavevmode\vadjust pre{\hypertarget{ref-vilesPlanningExperimentsFirst2008}{}}%
Viles, E., M. Tanco, L. Ilzarbe, and M. J. Alvarez. 2008. {``Planning
{Experiments}, the {First Real Task} in {Reaching} a {Goal}.''}
\emph{Quality Engineering} 21 (1): 44--51.
\url{https://doi.org/10.1080/08982110802425183}.

\leavevmode\vadjust pre{\hypertarget{ref-ggplot2}{}}%
Wickham, Hadley. 2016. \emph{{ggplot2}: Elegant Graphics for Data
Analysis}. Springer-Verlag New York.
\url{https://ggplot2.tidyverse.org}.

\leavevmode\vadjust pre{\hypertarget{ref-dplyr}{}}%
Wickham, Hadley, Romain François, Lionel Henry, and Kirill Müller. 2022.
\emph{{dplyr}: A Grammar of Data Manipulation}.
\url{https://CRAN.R-project.org/package=dplyr}.

\leavevmode\vadjust pre{\hypertarget{ref-Wilkinson2005-oz}{}}%
Wilkinson, Leland. 2005. \emph{The Grammar of Graphics}. 2nd ed.
Springer.

\leavevmode\vadjust pre{\hypertarget{ref-Winowiecki2011-zx}{}}%
Winowiecki, Leigh, Sean Smukler, Kenneth Shirley, Roseline Remans,
Gretchen Peltier, Erin Lothes, Elisabeth King, Liza Comita, Sandra
Baptista, and Leontine Alkema. 2011. {``Tools for Enhancing
Interdisciplinary Communication.''} \emph{Sustainability: Science
Practice and Policy} 7 (1): 74--80.
\url{https://doi.org/10.1080/15487733.2011.11908067}.

\leavevmode\vadjust pre{\hypertarget{ref-knitr}{}}%
Xie, Yihui. 2015. \emph{Dynamic Documents with {R} and Knitr}. 2nd ed.
Boca Raton, Florida: Chapman; Hall/CRC. \url{https://yihui.org/knitr/}.

\leavevmode\vadjust pre{\hypertarget{ref-rmarkdown}{}}%
Xie, Yihui, J. J. Allaire, and Garrett Grolemund. 2018. \emph{R
Markdown: The Definitive Guide}. Boca Raton, Florida: Chapman; Hall/CRC.
\url{https://bookdown.org/yihui/rmarkdown}.

\end{CSLReferences}

\end{document}